\documentclass[aps,prb,superscriptaddress,twocolumn,floatfix]{revtex4-1}
\usepackage{graphicx,color}
\usepackage{dcolumn}
\usepackage{bm}
\usepackage{stmaryrd}
\usepackage{latexsym}
\usepackage{amssymb}
\usepackage{amsfonts}
\usepackage{amsmath}
\usepackage{subfigure}
\usepackage{hyperref}
\usepackage{verbatim}
\usepackage{multirow} 

\begin{document}

\title{Extensibility of Hohenberg-Kohn Theorem to general quantum systems}

\author{Limin Xu}
\affiliation{Department of Mathematical Sciences, Tsinghua University, Beijing 100084, China}

\author{Jiahao Mao}
\affiliation{State Key Laboratory of Low Dimensional Quantum Physics and Department of Physics, Tsinghua University, Beijing 100084, China}

\author{Xingyu Gao}
\affiliation{Laboratory of Computational Physics, Institute of Applied Physics and Computational Mathematics, Beijing, 100088, China}

\author{Zheng Liu}
\email{E-mail: zheng-liu@tsinghua.edu.cn}
\affiliation{Institute for Advanced Study, Tsinghua University, Beijing 100084, China}

\begin{abstract}

Hohenberg-Kohn (HK) theorem is a cornerstone of modern electronic structure calculations. For interacting electrons, given that the internal part of the Hamiltonian ($\hat H_{int}$), containing the kinetic energy and Couloumb interaction of electrons, has a fixed form, the theorem states that when the electrons are subject to an external electrostatic field, the ground-state density can inversely determine the field, and thus the full Hamiltonian completely. For a general quantum system, a HK-type Hamiltonian in the form of $\hat H_{hk}\{g_i\}=\hat H_{int}+\sum_i g_i \hat O_i$ can always be defined, by grouping those terms with fixed or preknown coefficients into $\hat H_{int}$, and factorizing the remaining as superposition of a set of Hermitian operators $\{\hat O_i\}$. We ask whether the HK theorem can be extended, so that the ground-state expectation values of $\{\hat O_i\}$ as the generalized density can in principle be used as the fundamental variables determining all the properties of the system. We show that the question can be addressed by introducing the concept of generalized density correlation matrix (GDCM) defined with respect to the $\{\hat O_i\}$ operators. The invertibility of the GDCM represents a mathematically rigorous and practically useful criterion for the extension of HK theorem to be valid.  We apply this criterion to several representative systems, including the quantum Ising dimer, the frustration-free systems, N-level quantum systems with fixed inter-level transition amplitude and tunable level energies, and a fermionic Hubbard chain with inhomogeneous on-site interactions. We suggest that for a finite-size system, finding an invertible GDCM under one single $\{g_i\}$ configuration is typically sufficient to establish the generic extensibility of the HK theorem in the entire parameter space. 
\end{abstract}

\maketitle

\setlength{\parindent}{2em}

\section{Introduction}

One of the most powerful computational tools for condensed-matter physics, quantum chemistry, and even molecular biology is built within the framework of density functional theory (DFT)~\cite{RMP15DFT,ShamReview}. By utilizing electron density, rather than wavefunction, as the central variable, the complexity of the many-body problem is drastically reduced~\cite{KohnReview}. The Hohenberg-Kohn (HK) thereom~\cite{HK} provides a solid theoretical basis for the success of DFT, stating that the ground state (GS) electron density completely and uniquely determines the system, and thus all the properties of interacting electrons are in principle functionals of density.

The HK theorem has been extended in several different ways (See e.g. Refs.~\cite{Bauer83,SCDFT} and Sec. VIA of Ref.~\cite{KohnReview}). In the most general form, we can write the HK-type Hamiltonian as:
\begin{eqnarray} \label{Hhk}
\hat H_{hk}=\hat H_{int}+\hat H_{ext} \nonumber \\
\hat H_{ext}=\sum_{i=1}^{N} g_i \hat O_i,
\end{eqnarray}
where $\hat H_{int}$ represents the fixed intrinsic part of the system, and $\hat H_{ext}$ represents the unfixed part factorized by a set of Hermitian operators $\{\hat O_i\}$. $\textbf{g}\triangleq (g_1, g_2, ..., g_N)^T$ contains the superposition coefficients and $\hat H_{hk}$ is a function of $\mathbf{g}$. Heuristically, the generalized density is defined as $\textbf{O}\triangleq(O_1, O_2, ..., O_N)^T$, and
\begin{eqnarray}
O_i\triangleq\langle \hat O_i\rangle_{GS},
\end{eqnarray}
in which $\langle...\rangle_{GS}$ stands for the GS expectation value. The label $i$ represents some general index, not necessarily the spatial coordinate. If the index is continuous, the summation later on should be interpreted as integrals. We do not demand $\{\hat O_i\}$ mutually commute either. Therefore, for an arbitrary Hamiltonian $\hat H$, by grouping those terms with fixed or preknown coefficients into $\hat H_{int}$, and the remaining into $\hat H_{ext}$, we can always transform $\hat H$ into $\hat H_{hk}$. In the limiting case, where no parameter in $\hat H$ is fixed, $\hat H_{int}=\varnothing$, and $\hat H_{ext}=\hat H$. A direct parallel can be drawn back to the interacting electrons, where $\hat H_{int}$ corresponds to the electron kinetic energy plus Coulomb interactions, $g_i$ corresponds to the external electrostatic potential $V(\mathbf{r})$, and $\hat O_i$ corresponds to the electron density operator $\hat n(\mathbf{r})$. The original HK theorem was thus proved for a special case of $\hat H_{hk}$.

Does a HK-type theorem always exist for an arbitrary choice of $\hat H_{int}$ and $\{\hat O_i\}$? Namely, $\hat H_{hk}$ can in principle be inversely determined by $\textbf{O}$. The answer is not a simple ``yes''. For example, in Ref.~\cite{graphDFT}, Penz and van Leeuwen insightfully pointed out several misconceptions in formulating the lattice DFT in the discretized electron model. 

A primary goal of this Article is to provide a simple-to-use criterion to identify systems violating the HK theorem. We will start by discussing a toy counter-example - the quantum Ising dimer (Sec. \ref{isingdimer}). Then, in Sec. \ref{proof}, we introduce the concept of generalized density correlation matrix (GDCM) following a revisit of HK's original proof, and demonstrate that the invertibility of the GDCM is the necessary and sufficient condition for HK's \textit{reductio ad absurdum} to complete. We also draw a connection between the invertibility of GDCM and the response function. In Sec. \ref{app}, we apply the GDCM criterion to frustration-free systems, a N-level quantum system and the fermionic Hubbard model. Based on the numerical results and analytical arguments, we suggest that for a finite-size system, finding an invertible GDCM under one single $\{g_i\}$ configuration is typically sufficient to establish the generic extensibility of the HK theorem in the entire parameter space. Sec. \ref{conclusion} concludes this Ariticle.

The significance of expanding the territory of the HK theorem is twofold, as envisioned by Capelle and Campo Jr. in their Review entitled ``Density functionals and model Hamiltonians: Pillars of many-particle physics''~\cite{PhysRep13}. On one hand, for systems the theorem applies, the computational tools developed within the state-of-the-art DFT calculations can be adapted to reduce the complexity of traditional many-body methods, in particular for spatial inhomogeneous cases. On the other hand, model systems designed for demonstrating exotic quantum properties, such as quantum criticality, long-range entanglement and fractional exictations, can be used as the theoretical laboratory to deepen our understandings of DFT in regimes difficult to be accessed in real electronic systems. In practice, generalized DFT have been designed and tested in a variety of quantum systems, such as the quantum Ising chain~\cite{PRB21Mao,ShamEntangle}, the Heisenberg model~\cite{PRB07SpinDFT,PRB10SpinDFT}, strongly-correlated lattice models~\cite{Gunnarsson95,GrapheneDFT,AttractHubbardDFT,RepulsiveFermionDFT,LuttingerDFT,LatticeDFT} and nuclei~\cite{nuclearDFT}. The concept of GDCM proposed in this Article serves as a mathematically rigorous and practically useful justification of such applications, and renders an early recognition of inapplicable cases.

\section{A toy counter-example}\label{isingdimer}

The quantum Ising dimer represents a simple quantum system violating the HK theorem. The Hamiltonian is:
\begin{eqnarray}
\hat H_{int}&=&-\hat\sigma_1^x\hat\sigma_2^x\nonumber\\
\hat H_{ext}&=&-g_1\hat\sigma_1^z-g_2\hat\sigma_2^z, 
\end{eqnarray}
where $\hat\sigma^{z/x}$ is the familiar $z$/$x$-component of the Pauli matrices and the generalized density is 
\begin{eqnarray}
\sigma^z_i= \langle \hat \sigma^z_i\rangle_{GS}.
\end{eqnarray}
In the basis spanned by the $\hat \sigma_i^z$ eigenstates: $|\uparrow\uparrow\rangle$, $|\downarrow\downarrow\rangle$, $|\uparrow\downarrow\rangle$ and $|\downarrow\uparrow\rangle$, 
\begin{eqnarray}\label{eq:isingdimer}
\hat H_{hk}=\left(\begin{array}{cccc}
  -g_1-g_2 & -1 & 0 & 0   \\
  -1 & g_1+g_2 & 0 & 0 \\
  0 & 0 & -g_1+g_2 & -1\\
  0 & 0 & -1 & g_1-g_2
\end{array}\right).
\end{eqnarray}
Diagonalizing this $4\times4$ matrix gives:
\begin{eqnarray}\label{eq:sigma}
\sigma_1^z=\sigma_2^z=\frac{g_1+g_2}{\sqrt{1+(g_1+g_2)^2}} ~(g_1g_2\geqslant 0)\nonumber \\
\sigma_1^z=-\sigma_2^z=\frac{g_1-g_2}{\sqrt{1+(g_1-g_2)^2}} ~(g_1g_2\leqslant 0).
\end{eqnarray}
It is clear that $\{\sigma_1^z, \sigma_2^z\}$ cannot determine $\{g_1, g_2\}$ uniquely and completely. 

It is still possible to declare a weaker version of DFT applying for the GS energy, in analogy to Levy's universal variational functional~\cite{Levy79}. However, one have to be cautious that the missing of the one-to-one correspondence between $\{\sigma_1^z, \sigma_2^z\}$ and $\{g_1, g_2\}$ leads to a rather singular range of definition of the density function, which makes an unconstrained variation problematic. Specifically, based on Eqs. (\ref{eq:isingdimer}) and (\ref{eq:sigma}), the GS energy of this model can be written as:
\begin{eqnarray}\label{eq:Egs}
E_{GS}=-\sum_{i=1}^2(\frac{1}{2}\sqrt{1-(\sigma_i^z)^2}+g_i\sigma_i^z).
\end{eqnarray}
Therefore, a universal function $E_{int}[\sigma_i^z]=-\frac{1}{2}\sum_{i=1}^2\sqrt{1-(\sigma_i^z)^2}$ can indeed be defined, valid for any $\{g_1,g_2\}$, for the internal energy, but according to Eq. (\ref{eq:sigma}),  its range of definition is clearly not the whole $\sigma_{1,2}^z\in (-1,1)$ 2D square, but two 1D segments: $|\sigma_1^z|=|\sigma_2^z|\in (-1,1)$. This is a reminiscence of the long-studied ``$V$-representability'' problem of interacting electrons \cite{Vrep}. 

\section{Theoretical discussion}\label{proof}

Keeping this counter-example in mind, to extend the HK theorem to an arbitrary choice of $\hat H_{int}$ and $\{\hat O_i\}$, we are urged to revisit HK's original proof~\cite{HK} carefully.  

\subsection{A revisit of HK's original proof}
HK's proof employs a \textit{reductio ad absurdum} argument, which can be easily extended to the the general form of $\hat H_{hk}$. Suppose that there were two different $\textbf{g}^{(1)}$ and $\textbf{g}^{(2)}$ leading to the same $\textbf{O}$. Denote the GS's of $\hat H_{hk}^{(1,2)}=\hat H_{int}+\sum_{i=1}^{N} g_i^{(1,2)}\hat O_i$ as $| GS^{(1,2)}\rangle$ , respectively. We have
\begin{eqnarray}\label{inequal1}
\langle GS^{(1)} | \hat H_{hk} ^{(1)} | GS^{(1)}\rangle &\leqslant& \langle GS^{(2)} | \hat H_{hk} ^{(1)} | GS^{(2)}\rangle \nonumber \\
\Rightarrow \langle GS^{(1)} | \hat H_{int} | GS^{(1)}\rangle &\leqslant& \langle GS^{(2)} | \hat H_{int} | GS^{(2)}\rangle, 
\end{eqnarray}
and
\begin{eqnarray}\label{inequal2}
\langle GS^{(2)} | \hat H_{hk} ^{(2)} | GS^{(2)}\rangle &\leqslant& \langle GS^{(1)} | \hat H_{hk} ^{(2)} | GS^{(1)}\rangle \nonumber \\
\Rightarrow \langle GS^{(2)} | \hat H_{int} | GS^{(2)}\rangle &\leqslant& \langle GS^{(1)} | \hat H_{int} | GS^{(1)}\rangle.
\end{eqnarray}
The two inequalities cannot be satisfied simultaneously, unless $|GS^{(2)}\rangle$ ($|GS^{(1)}\rangle$) happens to also be the ground state of $H_{hk} ^{(1)}$ ($H_{hk} ^{(2)})$. 

Note that we cannot simply replace  ``$\leqslant$'' in (\ref{inequal1}) and (\ref{inequal2}) with ``$<$'' by demanding that the GS is non-degenerate, because it is still possible that $|GS^{(1)}\rangle=|GS^{(2)}\rangle$. Actually, GS degeneracy is irrelevant to the proof, as explicitly remarked by Kohn himself ~\cite{KohnReview}. Regardless the GS degeneracy, the equal sign can be removed if we further have the precondition that \textit{$\hat H_{hk}^{(1,2)}$ do not share a common ground state}. With this precondition, the two inequalities contradict, and the supposition that two different $\textbf{g}^{(1)}$ and $\textbf{g}^{(2)}$ can lead to the same $\textbf{O}$ has to be refuted. The \textit{reductio ad absurdum} then completes. For interacting electrons, this precondition is considered to be satisfied automatically except for $V^{(1)}(\mathbf{r})-V^{(2)}(\mathbf{r})=const$, but for the general cases, we should not take it for granted.

\subsection{GDCM}

Our problem is then reduced to exclude the possibility that two different $\textbf{g}^{(1)}$ and $\textbf{g}^{(2)}$ produce the same $|GS\rangle$. Crucially, when this possibility occurs,  $|GS\rangle$ will also be an eigenstate of $\hat H_{hk} ^{(1)}-\hat H_{hk} ^{(2)}=\sum_{i=1}^{N} \Delta g_i \hat O_i$, which can be equivalently expressed as 
\begin{eqnarray}\label{variance}
\langle (\sum_{i=1}^{N} \Delta g_i \hat O_i)^2 \rangle_{GS}-\langle \sum_{i=1}^{N} \Delta g_i \hat O_i \rangle^2_{GS}=0,
\end{eqnarray}
i.e. the variance or ``fluctuation'' of $\sum_{i=1}^{N} \Delta g_i \hat O_i$ with respect to $|GS\rangle$ is zero. Equation (\ref{variance}) can be transformed into a more compact form:
\begin{eqnarray}\label{quad}
\Delta\mathbf{g}^T \mathbf{M}\Delta\mathbf{g}=0,
\end{eqnarray}
in which $\mathbf{M}$ is a real symmetric matrix (an operator for the continuous indices) defined as
\begin{eqnarray}\label{M}
M_{ij}=\frac{1}{2}\langle \hat O_i \hat O_j + \hat O_j \hat O_i \rangle_{GS}-\langle\hat O_i\rangle_{GS}\langle\hat O_j\rangle_{GS},
\end{eqnarray}

When $\Delta\mathbf{g}\neq0$, Eq. (\ref{quad}) can be satisfied, iff the determinant of $\textbf{M}$ is zero and $\Delta\mathbf{g}$ is proportional to an eigenvector of $\textbf{M}$ with zero eigenvalue. Therefore, if we find that the $|GS\rangle$ produces an invertible $\textbf{M}$, i.e. zero is not an eigenvalue or $|\textbf{M}|\neq 0$,  it is guaranteed that this $|GS\rangle$ cannot be shared by two different 
$\textbf{g}$'s.  

The interesting point is that $M_{ij}$ physically represents the equal-time GS correlation between a pair of generalized density operators. Accordingly, we term $\textbf{M}$ as the GDCM, the invertibility of which determines the extensibility of the HK theorem. 

It is instructive to check the GDCM of the quantum Ising dimer at this point. From Eq. (\ref{eq:isingdimer}), the GS has the form:
\begin{eqnarray}
|GS\rangle=\begin{cases}
c_1 |\uparrow\uparrow\rangle + c_2 |\downarrow\downarrow \rangle ~(g_1g_2\geqslant 0)\nonumber \\
c_1 |\uparrow\downarrow\rangle + c_2 |\downarrow\uparrow \rangle ~(g_1g_2\leqslant 0),
\end{cases}
\end{eqnarray}
in which $c_{1,2}$ are coefficients dependent on $\{g_1,g_2\}$. The GDCM is:
\begin{eqnarray}
\textbf{M}=\left(\begin{array}{cc}
  1-(\sigma_1^z)^2 & sgn(\sigma_1^z\sigma_2^z)-\sigma_1^z\sigma_2^z    \\
  sgn(\sigma_1^z\sigma_2^z)-\sigma_1^z\sigma_2^z & 1-(\sigma_2^z)^2
\end{array}\right),
\end{eqnarray}
in which $sgn$ is the sign function. With $\sigma_{1,2}^z$ given in Eq. (\ref{eq:sigma}), $\det (M)\equiv0$, signifying the complete absence of the one-to-one correspondence between $(g_1, g_2)$ and $(\sigma_1^z, \sigma_2^z)$.

\subsection{Alternative perspective from the response function}\label{rf}

Besides the GDCM, the possibility that two different $\textbf{g}^{(1)}$ and $\textbf{g}^{(2)}$ produce the same $|GS\rangle$ can be alternatively dictated by the response function. If $|GS\rangle$ is an eigenstate of $\hat H_{hk} ^{(1)}-\hat H_{hk} ^{(2)}=\sum_{i=1}^{N} \Delta g_i \hat O_i$, we can define a perturbing operator proportional to $\sum_{i=1}^{N} \Delta g_i \hat O_i$ that does not generate any response to the GS. This requires that the static response function $\chi_{ij}=\frac{\partial O_i}{\partial g_j}$ is noninvertible. Relatedly, the invertibility of dynamical response function has long been discussed in the context of time-dependent DFT~\cite{TDDFT84,tddftreview,RFInv}. We note that although $\chi_{ij}$ can be viewed as the Jacobian matrix of the map from \textbf{g} to \textbf{O}, the HK theorem is not a trivial application of the inverse function theorem, which asserts local invertibility only. To establish the global one-to-one correspondence between \textbf{g} and \textbf{O}, it requires the GS condition \footnote{The lack of HK theorem for excited states was discussed in \cite{exHK}} and further proof as elaborated in Secs. 3.1 and 3.2.

An advantage of the GDCM is that given the GS at a single $\mathbf{g}$,  the GDCM at this point can be immediately calculated, whereas to calculate the response function requires additional information. This makes the GDCM a more convenient criterion to use, as we will show in the next section.

\section{Applications}\label{app}
\subsection{Frustration-free systems}
Following the terminology used in quantum information (See e.g. Sec. 4.5 in Ref.~\cite{ZengQuantum}), We call $\hat H_{hk}$ frustration-free, if the GS of $\hat H_{hk}$ is also the GS of $\hat H_{int}$ and each $\hat O_i$. It is straightforward to show that the GDCM is 0. Therefore, the HK theorem does not apply. 

It is worth mentioning that such seemingly classical systems play an important role in demonstrating key concepts in quantum computation, as well as unconventional quantum states of matters. Well-known examples include Kitaev's toric code model~\cite{toriccode} and the Affleck-Kennedy-Lieb-Tasaki spin model~\cite{AKLT}. Regretfully, insights into generalized DFT reflecting these novel features have to be sought from more complicated models containing frustrations.

\subsection{$N$-level quantum systems}\label{nlevel}
Consider a $N$-level quantum system with fixed inter-level transition amplitude and tunable level energies:
\begin{eqnarray}\label{kagomeHint}
\hat H_{int}&=&\sum_{i{\sim}j} |i\rangle\langle j|+ H.c.\label{hop} \\
\hat H_{ext}&=&\sum_{i=1}^N g_i |i\rangle\langle i|, 
\end{eqnarray}
in which $i{\sim}j$ denotes a pair of transition-allowed levels. The density has the conventional definition:
\begin{eqnarray}
n_i=\langle GS|i\rangle\langle i|GS\rangle. 
\end{eqnarray}
The GDCM can be calculated to be:
\begin{eqnarray}\label{kagomeM}
M_{ij}=n_i\delta_{ij}-n_in_j.
\end{eqnarray}
Its eigenvectors with zero eigenvalue consist of two types of vectors. Denoting the $N$ components of an eigenvector as $v_{1,...,N}$, one type is $v_{1,...,N}=1$, which trivially means that a constant shift of $g_i$ does not change the GS. The other less trivial type is $v_i=0$ if $n_i\neq 0$, and $v_i$ is an arbitrary number if $n_i=0$, which indicates that whenever the GS density vanishes at a level $i$, it fails to determine the level energy $g_i$. Therefore, the HK theorem will be violated by zero points in the GS density.

\begin{figure}
\centering
\includegraphics[width=9cm]{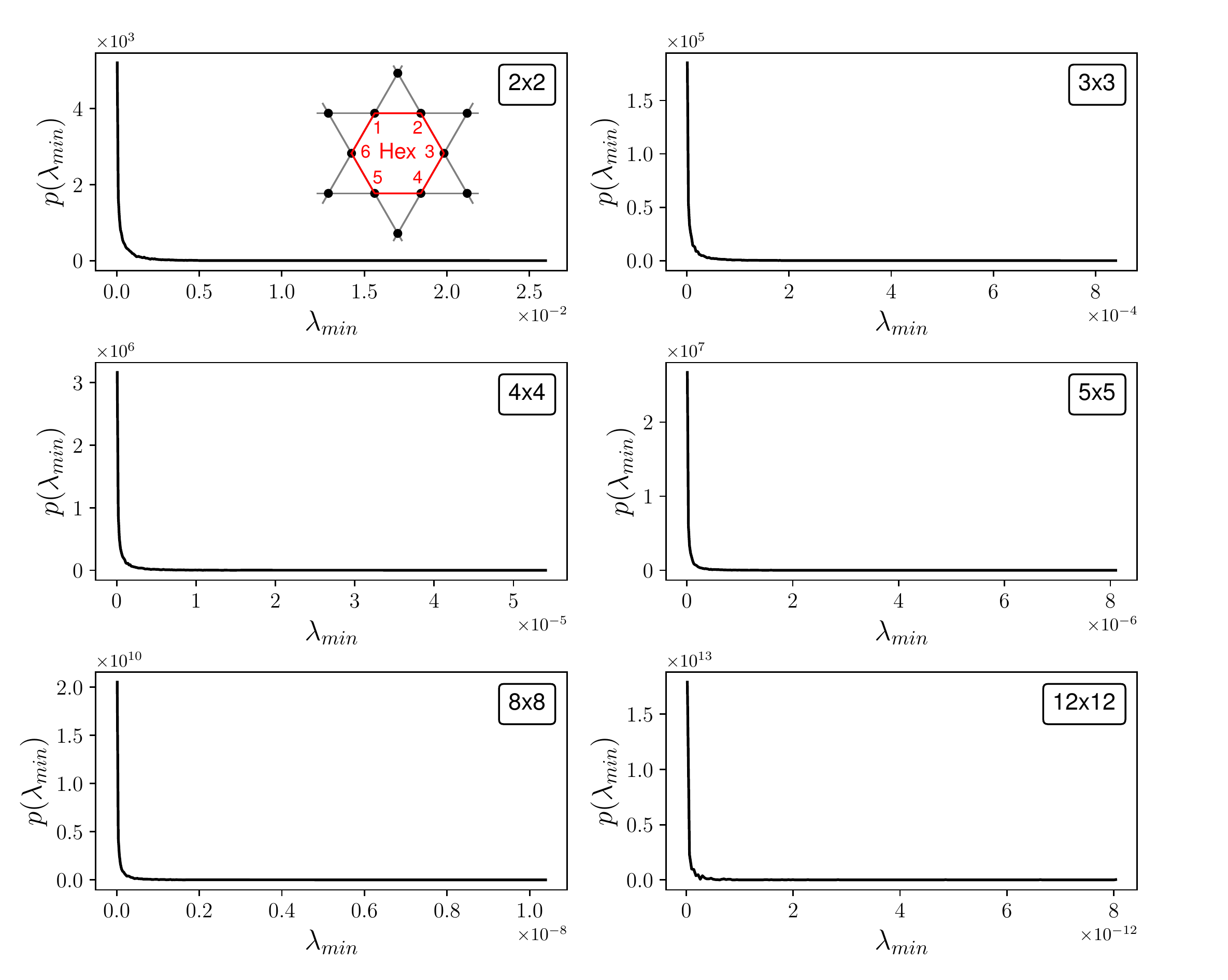}
\caption{Probability density distribution of the smallest eigenvalue $p(\lambda_{min}$) of the GDCM of a N-level quantum system when the HK theorem is invalid. The data is generated based on Eqs. (\ref{kagomeHint})-(\ref{kagomeM}), and by choosing $i{\sim}j$ as a pair of nearest-neighbor vertices of a finite kagome lattice (inset) with periodic boundary conditions. The legend of each figure denotes the size of the lattice in terms of the lattice vectors. $\textbf{g}$ is randomly generated according to a uniform distribution between -1 and 1. $p(\lambda_{min})$ is evaluated by statistically counting the frequency of $\lambda_{min}$ falling in a small interval, and normalized by $\int p(\lambda)d\lambda=1$. The zero eigenvalue associated with a trivial constant shift of the energy levels is excluded.}
\label{kagome}
\end{figure}

This conclusion does not rely on the detailed structure and the exact strength of the transitions. To show a GS with extensive zero points, we choose $i{\sim}j$ as a pair of nearest-neighbor vertices of a kagome lattice, where the destructive interference of multiple transition paths due to the special geometry is known to produce compact localized eigenstates~\cite{CPBFlat}.  It is also important to note that the transition amplitude is purposely set to be positive in Eq. (\ref{kagomeHint}). It is straightforward to check that $|\psi_{L}\rangle=\frac{1}{\sqrt 6}\sum_{i\in \text{Hex}} (-1)^i |i\rangle$ is a GS of $\hat H_{int}$. The summation index $i$ runs around the vertices of a hexagon (Hex) as marked in the inset of Fig. \ref{kagome}. The associated ground state density is $n_{i\in \text{Hex}}=\frac{1}{6}$; $n_{i\notin \text{Hex}}=0$. By keeping $g_{i\in \text{Hex}}=0$ and raising $g_{i\notin \text{Hex}}$, $|\psi_{L}\rangle$ can still be the ground state of a large set of $\mathbf{g}$, which corroborates the GDCM result.

\begin{figure}
\centering
\includegraphics[width=9cm]{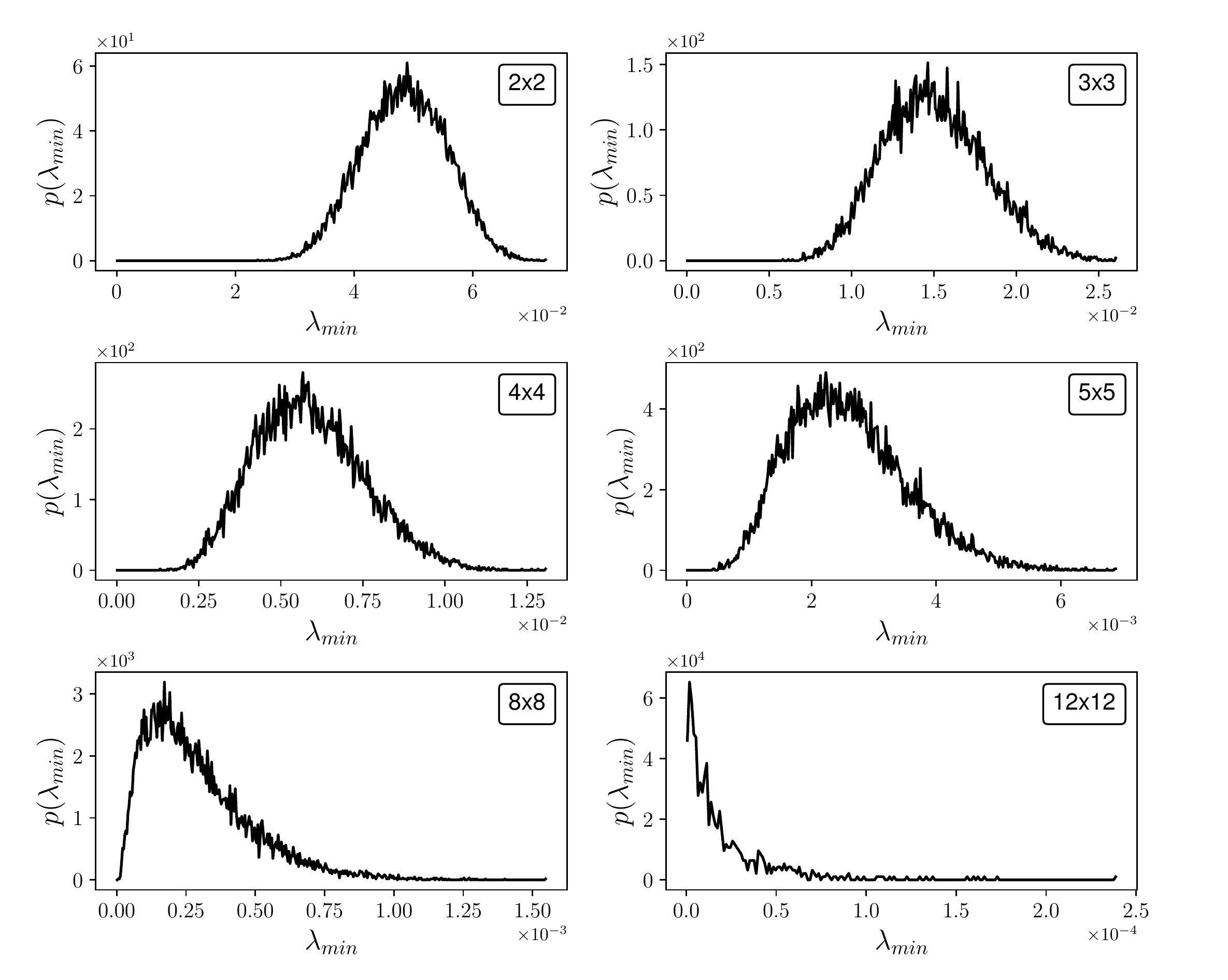}
\caption{Probability density distribution of the smallest eigenvalue $p(\lambda_{min}$) of the GDCM of a N-level quantum system when the HK theorem is valid. The data is generated in the same way as described in the caption of Fig. \ref{kagome}, except for a negative sign added to Eq. (\ref{kagomeHint}).}
\label{nkagome}
\end{figure}

As a numerical justification, we evaluate the GDCM under randomly generated $\{g_i\}$. The smallest eigenvalue $\lambda_{min}$ of $\textbf{M}$ (excluding the trivial zero associated with the constant shift of $\mathbf{g}$) is statistically analyzed (Fig. \ref{kagome}). The distribution shows a predominating peak at a vanishing $\lambda_{min}$, indicating generic noninvertibility of $\textbf{M}$. This is expected to be a typical behavior for HK-theorem invalid systems.
 
 For comparison, we redo the statistical analysis after inverting the transition amplitude from +1 to -1 (Fig. \ref{nkagome}), which makes $\hat H_{int}$ equivalent to a discretized kinetic energy operator. As elegantly discussed in Ref.~\cite{graphDFT} in the context of a free electron hopping model, for a connected graph, $n_i$ is strictly positive based on the Perron-Frobenius theorem from the linear algebra. Therefore, the GDCM is protected to be invertible. The $\lambda_{min}$ distribution becomes sharply different from Fig. \ref{kagome}. In particular, we see that it shows a clear peak at a finite $\lambda_{min}$, and quickly decays when $\lambda_{min}$ approaches zero. This is expected to be a typical behavior for HK-theorem valid systems. 
 
$\lambda_{min}$ serves as a useful measure of the sensitivity to errors of the inversion from the GS density to $\hat H_{hk}$. A nonzero but small $\lambda_{min}$ implies that although the GS density determines $\mathbf{g}$ uniquely and completely, two different $\mathbf{g}$'s can result in very similar GS density. It is interesting to notice that when the system size increases, the most probable $\lambda_{min}$ keeps shifting to smaller values, signifying an overall more ill-conditioned inversion. This evolution should be attributed to Anderson localization generated by the random $\mathbf{g}$~\cite{AndersonTrans}. When the system size is larger than the localization length, $n_i$ becomes exponentially small away from the localization center. 

\subsection{Fermionic Hubbard model}\label{hubbard}
Consider that the Fermion Hubbard model contains inhomogeneous onsite interactions:
\begin{eqnarray}\label{eq:hubbard}
\hat H_{int}&=&-\sum_{i{\sim}j,\sigma}\hat c_{i\sigma}^\dag \hat c_{j\sigma} + H.c.\nonumber \\
\hat H_{ext}&=&\sum_{i=1}^N g_i \hat c_{i\uparrow}^\dag \hat c_{i\uparrow} \hat c_{i\downarrow}^\dag \hat c_{i\downarrow},
\end{eqnarray}
in which $\hat c_{i\sigma}^\dag$ and $\hat c_{i\sigma}$ are the fermion creation and destruction operators respectively at a site $i$, and $\sigma=\{\uparrow,\downarrow\}$ denotes the spin degree of freedom. The meaning of $i{\sim}j$ is the same as in Sec.~\ref{nlevel}.

The generalized density is
\begin{eqnarray}
O_i&=&\langle \hat c_{i\uparrow}^\dag \hat c_{i\uparrow} \hat c_{i\downarrow}^\dag \hat c_{i\downarrow}\rangle_{GS} \nonumber \\
&=&\frac{1}{4}(\langle \hat n_i^2\rangle_{GS}-\langle \hat S_{z,i}^2\rangle_{GS}), 
\end{eqnarray}
physically characterizing the difference between on-site density fluctuation and spin fluctuation, with
\begin{eqnarray}
\hat n_i&\triangleq& \hat c_{i\uparrow}^\dag \hat c_{i\uparrow}+\hat c_{i\downarrow}^\dag \hat c_{i\downarrow} \\ \nonumber
\hat S_{z,i}&\triangleq& \hat c_{i\uparrow}^\dag \hat c_{i\uparrow}-\hat c_{i\downarrow}^\dag \hat c_{i\downarrow}.
\end{eqnarray}
Note that it is the generalized density $O_i$ instead of the conventional density $n_i$ that plays the central role here. 

For a many-body model, it is computationally much more demanding to make a statistical plot of $\lambda_{min}$ like Figs. \ref{kagome} and \ref{nkagome} by sampling the $\mathbf{g}$ space. Is there a convenient way to determine whether the HK theorem generically applies or not? An enlightening observation of Figs. \ref{kagome} and \ref{nkagome} is that a ``one-for-all'' test should typically work: given the GDCM at a single $\mathbf{g}$, if it is found to be invertible, the extensibility of the HK theorem is expected to be generically valid in the entire parameter space. 

This observation can be rationalized for finite-size systems, where the GS is typically an analytical function of $\textbf{g}$~\footnote{Formal discussions on the GS analyticity under the nondegeneracy condition can be found in Sec. IV of Ref.~\cite{graphDFT} by quoting the Rellich theorem. For a pedagogical explanation of the Rellich theorem, we refer to Chpt. Introduction of Ref.~\cite{kato1995perturbation}). }. Then, $|\mathbf{M}|$ is also an analytical function of $\textbf{g}$, considering that the determinant of the GDCM is a multivariable polynomial about the coefficients of the GS. A useful property of the analytic function is that its zero set has a zero measure, unless the function is identically zero. Accordingly, either $|\textbf{M}|\equiv 0$, or $|\textbf{M}|\not\equiv 0$ except for a measure zero set of $\textbf{g}$ points, as reflected in Figs. \ref{kagome} and \ref{nkagome}. 

For the Hubbard model, the most convenient point to check is $\textbf{g}=0$, where the GDCM is calculable via Wick's theorem. Let us consider a Hubbard chain with the periodic boundary condition. Without interactions, the GS is easily solved by performing Fourier transformation. To have a unique GS, we can declare that  $N~\text{mod}~4=2$ and the total fermion number equals $N$. Restricting the discussion to such half-filled chains is only for the purpose of demonstration simplicity. Evaluating the noninteracting GDCM in higher dimensional lattices and/or under an arbitrary filling is always feasible, which certainly deserves a more systematic calculation.

For the Hubbard chain defined above, with some derivations, the GDCM at $\textbf{g}=0$ has the coefficient:
\begin{equation}\label{M_hubbard}
M_{ij}=\begin{cases}
\frac{3}{16}, & i=j\\
\left[\frac{1}{4}-\frac{1-\left(-1\right)^{\left(i-j\right)}}{2N^{2}\sin^{2}\left(\frac{\pi}{N}\left(i-j\right)\right)}\right]^{2}-\frac{1}{16}, & i\neq j.
\end{cases}
\end{equation}
Its eigenvalues are 
\begin{eqnarray}\label{lambda_hubbard}
&\lambda(p)= \frac{3}{16}+\nonumber \\ 
&\frac{1}{16} \sum_{j=1}^{N-1}\left(\left[1-\frac{2\left(1-\left(-1\right)^{j}\right)}{N^{2}\sin^{2}\left(\frac{\pi j}{N}\right)}\right]^{2}-1\right)\cos\left(jp\right),
\end{eqnarray}
with $p\in\left\{ 0,\frac{2\pi}{N},\dots,\frac{2\pi}{N}\left(N-1\right)\right\}$, which can be proved to be strictly positive. We include details of the proof in the Appendix. The point is that so long as $N$ is finite, we can immediately infer that the HK theorem is generically valid. 

\section{Conclusion and prospect}\label{conclusion}
In conclusion, we propose the GDCM as a key quantity to examine the extensibility of the HK theorem to general quantum systems. The ``one-for-all'' test we used for the Hubbard chain should find general applications for finite-size quantum systems. Going to the thermodynamic limit demands more cares, because phase transitions break analyticity. One should at least check one point for each phase and avoid phase boundaries. Numerically, an extrapolation from the finite-size results to the thermodynamic limit is possible, but should be performed carefully (See e.g. Sec. IX in \cite{RMP01QMC} in the context of quantum Monte Carlo).

Our dicussion bridges the HK theorem to a fundamental question in quantum information - what is the minimal information required to fully determine a quantum system? The correlation matrix was first introduced in Ref. \cite{Qi2019} to inversely reconstruct a local Hamiltonian. In Ref. \cite{PRX18}, the possibility of a single eigenstate to encode the full Hamiltonian is conceptually related to the eigenstate thermalization hypothesis. Recalling that an extended HK theorem is rationalized by excluding the possibility that two $\hat H_{hk}$'s produce the same GS, we can surely sense some deep connections. Indeed, in every counter-example we showed violating the HK theorem, certain sorts of ``localization'' of the GS can be recognized. In prospect, the emerging topics in quantum technologies, such as many-body localization, entanglement entropy and topological orders, are expected to shed refreshing light on DFT.

\section{Acknowledgements} \par 
We would like to thank Hui Zhai for bringing Ref.~\cite{Qi2019} to our attention, Xiaoliang Qi for elaborating the concept of correlation matrix developed in his paper, and Duanyang Liu for critically reading the manuscript. Research supported by Tsinghua University Initiative Scientific Research Program. L.X. and J.M. contributed equally to this work. 

\appendix
\section{Invertibility of the GDCM of a half-filled Hubbard chain at the noninteracting limit}
We rewrite the Hamiltonian of Eq. (\ref{eq:hubbard}) in 1D:
\begin{eqnarray}
\hat H_{int}&=&-\sum_{n=1}^N\hat c_n^\dag \hat c_{n+1} + H.c.\nonumber \\
\hat H_{ext}&=&\sum_{n=1}^N g_n \hat c_{n\uparrow}^\dag \hat c_{n\uparrow} \hat c_{n\downarrow}^\dag \hat c_{n\downarrow},
\end{eqnarray}
and consider the specific case $g_n=0$, $N~\text{mod}~4=2$ with the periodic boundary condition. At half filling, the fermion number equals $N$.

The generalized density is
\begin{eqnarray}
O_n&=&\langle \hat c_{n\uparrow}^\dag \hat c_{n\uparrow} \hat c_{n\downarrow}^\dag \hat c_{n\downarrow}\rangle_{GS}.
\end{eqnarray}

By defining the the Fourier transformation as:
\begin{align}
\hat c_{n\sigma} & =\frac{1}{\sqrt{N}}\sum_{k}e^{\mathrm{i}kn}\hat c_{k\sigma}\\
\hat c_{n\sigma}^{\dagger} & =\frac{1}{\sqrt{N}}\sum_{k}e^{-\mathrm{i}kn}\hat c_{k\sigma}^{\dagger},
\end{align}
with $k\in\left\{-\pi, -\pi+\frac{2\pi}{N},\dots,\pi-\frac{2\pi}{N}\right\}$, the ground state is:
\begin{equation}
|\text{GS}\rangle=\prod_{-\frac{\pi}{2}<k<\frac{\pi}{2}}\prod_{\sigma=\uparrow,\downarrow}\hat c_{k\sigma}^{\dagger}|0\rangle.
\end{equation}

The two terms in the GDCM can be readily calculated:
\begin{align}
\left\langle \hat{O}_{n}\right\rangle  & =\left\langle c_{n\uparrow}^{\dagger}c_{n\uparrow}c_{n\downarrow}^{\dagger}c_{n\downarrow}\right\rangle \nonumber \\
 & =\frac{1}{N^{2}}\sum_{k_{1}k_{2}k_{3}k_{4}}e^{\mathrm{i}\left(-k_{1}+k_{2}-k_{3}+k_{4}\right)n}\left\langle c_{k_{1}\uparrow}^{\dagger}c_{k_{2}\uparrow}c_{k_{3}\downarrow}^{\dagger}c_{k_{4}\downarrow}\right\rangle \nonumber \\
 & =\frac{1}{N^{2}}\sum_{\left|k\right|<\frac{\pi}{2}}\sum_{\left|k'\right|<\frac{\pi}{2}}1=\frac{1}{4}
\end{align}

\begin{align}
\left\langle \hat{O}_{n}\hat{O}_{m}\right\rangle = & \left\langle c_{n\uparrow}^{\dagger}c_{n\uparrow}c_{n\downarrow}^{\dagger}c_{n\downarrow}c_{m\uparrow}^{\dagger}c_{m\uparrow}c_{m\downarrow}^{\dagger}c_{m\downarrow}\right\rangle \nonumber \\
= & \frac{1}{N^{4}}\left[\sum_{\left|k_{1}\right|<\frac{\pi}{2}}\sum_{\left|k_{2}\right|<\frac{\pi}{2}}1+\sum_{\left|k_{1}\right|<\frac{\pi}{2}}\sum_{\left|k_{2}\right|>\frac{\pi}{2}}e^{-\mathrm{i}\left(k_{1}-k_{2}\right)\left(n-m\right)}\right]^{2}\nonumber \\
= & \begin{cases}
\frac{1}{4}, & n=m\\
\left[\frac{1}{4}-\frac{1-\left(-1\right)^{\left(n-m\right)}}{2N^{2}\sin^{2}\left(\frac{\pi}{N}\left(n-m\right)\right)}\right]^{2}, & n\neq m
\end{cases}
\end{align}

Thus
\begin{equation}
M_{nm}=\begin{cases}
\frac{3}{16}, & n=m\\
\left[\frac{1}{4}-\frac{1-\left(-1\right)^{\left(n-m\right)}}{2N^{2}\sin^{2}\left(\frac{\pi}{N}\left(n-m\right)\right)}\right]^{2}-\frac{1}{16}, & n\neq m
\end{cases}
\end{equation}

Because $\textbf{M}$ has translational symmetry, we can diagonalize it with Fourier transformation. The eigenvalues have the property:
\begin{align*}
\lambda\left(p\right) & =\frac{3}{16}+\frac{1}{16}\sum_{n=1}^{N-1}\left(\left[1-\frac{2\left(1-\left(-1\right)^{n}\right)}{N^{2}\sin^{2}\left(\frac{\pi n}{N}\right)}\right]^{2}-1\right)\cos\left(np\right)\\
 & >\frac{3}{16}-\frac{1}{16}\sum_{n=1}^{N-1}\left|\left[1-\frac{2\left(1-\left(-1\right)^{n}\right)}{N^{2}\sin^{2}\left(\frac{\pi n}{N}\right)}\right]^{2}-1\right|\left|\cos\left(np\right)\right|\\
 & \ge\frac{3}{16}-\frac{1}{16}\sum_{n=1}^{N-1}\left|\left[1-\frac{2\left(1-\left(-1\right)^{n}\right)}{N^{2}\sin^{2}\left(\frac{\pi n}{N}\right)}\right]^{2}-1\right|\\
 & =\frac{3}{16}+\frac{1}{16}\sum_{n=1}^{N-1}\left(\left[1-\frac{2\left(1-\left(-1\right)^{n}\right)}{N^{2}\sin^{2}\left(\frac{\pi n}{N}\right)}\right]^{2}-1\right),\\
\end{align*}
with $p\in\left\{ 0,\frac{2\pi}{N},\dots,\frac{2\pi}{N}\left(N-1\right)\right\}$.

In order to prove $\lambda(p)>0$, we just need to prove $\sum_{n=1}^{N-1}\left(\left[1-\frac{2\left(1-\left(-1\right)^{n}\right)}{N^{2}\sin^{2}\left(\frac{\pi n}{N}\right)}\right]^{2}-1\right)>-3$.
Since $N$ is even, we can divide the summation into two halves: 
\begin{equation}
    \begin{aligned}
   &\sum_{n=1}^{\frac{N}{2}}\left(\left[1-\frac{2\left(1-\left(-1\right)^{n}\right)}{N^{2}\sin^{2}\left(\frac{\pi n}{N}\right)}\right]^{2}-1\right)\\
   \ge &\sum_{n=1}^{\frac{N}{2}}\left(\left[1-\frac{2\left(1-\left(-1\right)^{n}\right)}{4n^2}\right]^{2}-1\right)\\
   =&\sum_{n=1,\mathrm{odd}}^{\frac{N}{2}} \left(\left[1-\frac{1}{n^2}\right]^{2}-1\right),
    \end{aligned}
\end{equation}
and
\begin{eqnarray}
\begin{aligned}
   &\sum_{n=\frac{N}{2}+1}^{N-1}\left(\left[1-\frac{2\left(1-\left(-1\right)^{n}\right)}{N^{2}\sin^{2}\left(\frac{\pi n}{N}\right)}\right]^{2}-1\right)\\
    \stackrel{m=N-n}{=}&\sum_{m=1}^{\frac{N}{2}-1}\left(\left[1-\frac{2\left(1-\left(-1\right)^{N-m}\right)}{N^{2}\sin^{2}\left(\frac{\pi m}{N}\right)}\right]^{2}-1\right)\\
     \ge&\sum_{m=1,\mathrm{odd}}^{\frac{N}{2}-1} \left(\left[1-\frac{1}{m^2}\right]^{2}-1\right),
\end{aligned}
\end{eqnarray}
where we use the fact: $\sin x\ge\frac{2}{\pi}x$, for $0<x\le\frac{\pi}{2}$.
Therefore, we have
\begin{equation}
    \begin{aligned}
       &\sum_{n=1}^{N-1}\left(\left[1-\frac{2\left(1-\left(-1\right)^{n}\right)}{N^{2}\sin^{2}\left(\frac{\pi n}{N}\right)}\right]^{2}-1\right)\\
       \ge&\sum_{n=1,\mathrm{odd}}^{\frac{N}{2}} \left(\left[1-\frac{1}{n^2}\right]^{2}-1\right)+\sum_{n=1,\mathrm{odd}}^{\frac{N}{2}-1} \left(\left[1-\frac{1}{n^2}\right]^{2}-1\right)\\
       \ge& 2\sum_{n=1,\mathrm{odd}}^{\infty} \left(\left[1-\frac{1}{n^2}\right]^{2}-1\right)\\
       =&2\sum_{n=1,\mathrm{odd}}^{\infty}\left[\frac{1}{n^4}-\frac{2}{n^2}\right]=2\cdot\frac{\pi^4}{96}-4\cdot\frac{\pi^2}{8}=-2.9054>-3.
    \end{aligned}
\end{equation}
The second inequality exploits the decreasing behavior of the summation with respect to $N$.\\
 \\
 \\ 
  \\
   \\
   \\
   \\
   \\
   \\
   \\
   \\
\medskip

\bibliography{refs.bib}

\end{document}